\documentclass[runningheads]{llncs}
\usepackage[T1]{fontenc}
\usepackage{graphicx}
\usepackage{orcidlink}
\usepackage{multirow}
\usepackage{tabularx}
\usepackage{booktabs}
\usepackage{array}
\usepackage{amsmath}
\usepackage{xurl}
\usepackage{fancyhdr}
\usepackage{makecell}

\begin{document}
\title{Leveraging Large Language Models for Automated Scalable Development of Open Scientific Databases }
%
%


\author{Nikita Gautam\inst{1}\orcidlink{0000-0002-1572-5405} \and
Doina Caragea\inst{1}\orcidlink{0000-0002-1572-5405} \and
Ignacio Ciampitti\inst{2}\orcidlink{0000-0001-9619-5129}
\and 
Federico Gomez\inst{2}\orcidlink{} 
} 
 \authorrunning{N. Gautam et al.}

\institute{ Department of Computer Science, Kansas State University, Manhattan, KS 66502, USA \and
Department of Agronomy, Purdue University, West Lafayette, IN 47907, USA 
}
\maketitle              

\setcounter{footnote}{0}
\begin{abstract}
With the exponential increase in online scientific literature, identifying reliable domain-specific data has become increasingly important but also very challenging. Manual data collection and filtering for domain-specific scientific literature is not only time-consuming but also labor-intensive and prone to errors and inconsistencies. To facilitate automated data collection, the paper introduces a web-based tool that leverages Large Language Models (LLMs) for automated and scalable development of open scientific databases. More specifically, the tool is based on an automated and unified framework that combines keyword-based querying, API-enabled data retrieval, and LLM-powered text classification to construct domain-specific scientific databases. Data is collected from multiple reliable data sources and search engines using a parallel querying technique to construct a combined unified dataset. The dataset is subsequently filtered using LLMs queried with prompts tailored for each keyword-based query to extract the relevant data to a scientific query of interest. The approach was tested across a set of variable keyword-based searches for different domain-specific tasks related to agriculture and crop yield. The results and analysis show 90\% overlap with small domain expert-curated databases, suggesting that the proposed tool can be used to significantly reduce manual workload. Furthermore, the proposed framework is both scalable and domain-agnostic and can be applied across diverse fields for building scalable open scientific databases. 

\end{abstract}

\section{Introduction}

Access to precise and organized scientific information remains a significant challenge in modern research. While new knowledge is generated at an accelerating pace each year, researchers often face difficulties retrieving and assembling information scattered across vast, distributed, and often not fully open-access sources to answer specific scientific questions. The fragmented nature of scientific information makes it labor-intensive to construct databases tailored to particular research needs. Researchers must manually search, extract, and curate information from multiple sources, an effort that is not only time-consuming but also prone to errors and inconsistencies. Prior studies in the agriculture field have highlighted persistent difficulties in retrieving and assembling data from existing literature, demonstrating gaps in data accessibility and standardization across various scientific domains \cite{ignacio1,ignacio2,ignacio3}. Systematic reviews and meta-analyses in applied plant science and agriculture, for example, often uncover inconsistencies and incomplete reporting that complicate the synthesis of reliable conclusions \cite{ignacio1,ignacio2,ignacio3}. Persistent challenges such as fragmented literature and limited coverage in areas like soil organic carbon (SOC) \cite{ledo2019global,beillouin2022global} and nutrient management underscore the need for scalable tools that help in data assembly. More specifically, there is a big need for practical, automated tools to retrieve information from distributed sources and assemble consistent and comprehensive, query-specific datasets to address targeted scientific questions.

To address this need, this paper introduces an automated web-based tool to enable scalable literature search, classification, and integration. The proposed tool combines keyword-based queries with zero-shot text classification leveraged by Large Language Models (LLMs) to identify and filter relevant publications efficiently from multiple distributed sources. By automating tasks that traditionally require extensive manual effort, such as screening articles for relevance, merging data from multiple sources, and removing irrelevant duplicates, the proposed tool significantly reduces the workload involved in assembling consistent, query-specific scientific datasets. This approach helps researchers efficiently build custom databases to answer targeted scientific questions across different domains. While this approach is general, we evaluate it on agriculture-specific tasks to highlight both its utility and domain challenges. 

The proposed tool integrates major scientific databases such as Scopus \cite{scopus2025}, Web of Science \cite{webofscience2025}, Science Direct \cite{sciencedirect2025}, and Google Scholar \cite{googlescholar2025}. It uses predefined keywords related to the specific research objective to guide the search across these sources. Retrieved records are then merged, and duplicates are identified and removed by matching Digital Object Identifiers (DOIs). Furthermore, article titles are used as an additional check when DOIs are missing or inconsistent. Only English-language publications are retained for further processing.

The tool employs LLMs to automate relevance classification and reduce bias and effort associated with manual filtering. 
Smaller LLMs such as, BERT \cite{bert} models can be used for training on domain-specific article titles to classify publications as relevant or irrelevant. While effective for the specific topics it is trained on, BERT-based classifiers require retraining to handle new subjects, limiting their general applicability. Moreover, in cases where data related to certain keywords are unavailable, training such models becomes infeasible altogether.
To overcome these limitations, more flexible LLMs such as ChatGPT \cite{gpt3}, LLaMA \cite{llama1}, and Gemini \cite{gemini} were used for prompt-based zero-shot classification, allowing the tool to adapt to diverse topics without retraining and improving its scalability across scientific fields.

The resulting tool can be used to compile structured, query-specific databases. Each resulting database includes key information such as paper titles, authors, publication years, DOIs, URLs, and more. To illustrate its application, we present a case study focused on an open database related to the yield response of major crops in Senegal to nutrient application. By systematically identifying, collecting, and organizing relevant publications, the tool supported the generation of yield response curves and the evaluation of predictors for nutrient impacts on crop yield. Preliminary tests demonstrated that the tool can significantly reduce manual effort. The framework is scalable and domain-agnostic, making it applicable for constructing open scientific databases in diverse research domains. To summarize, the key contributions of this work are as follows:

\begin{itemize}
    \item We design a pipeline for collecting and filtering large-scale scientific data with minimal supervision.
    \item We introduce an abstract filtering tool that integrates  LLM-based classification to automatically remove noisy data.
    \item We perform a comprehensive evaluation of our proposed approach to quantify the performance. 
    
\end{itemize}

\section{Background}

We review two main lines of work relevant to the proposed approach and evaluation, specifically, challenges posed by the development of topic-specific databases in agriculture and the use of language models to help in building such databases.
\subsection{Challenges in Assembling Agricultural Scientific Data}
Recent studies in the field of agriculture have highlighted the growing importance of soil organic carbon (SOC) in climate and agricultural research \cite{ignacio3}. Large-scale compilations of SOC data at global and regional levels have been developed \cite{ledo2019global,beillouin2022global}. However, these efforts still reveal significant gaps in accessible, comprehensive SOC datasets, especially at finer regional scales. Similarly, recent reviews have addressed the need for site-specific nutrient management strategies by compiling data on crop yield responses to nutrient applications into open-access databases \cite{ignacio1}. Such compilations aim to improve data availability for major field crops and support better analysis and decision-making in agricultural management. While these domain-specific initiatives demonstrate the potential of open-access databases to advance research \cite{source1}, they also underscore persistent challenges in creating and maintaining comprehensive, updated datasets. Efforts to establish robust open data infrastructures remain limited in many fields \cite{source2}, and practical barriers such as data ownership, privacy, and stakeholder interests complicate collaborative data sharing \cite{source3}. Although some structured databases exist for some topics \cite{open1,open2,open3}, they are often narrowly scoped, require significant expert input, and are difficult to scale or adapt for new research questions. In addition, many current approaches lack automation and standardization, making it labor-intensive to collect and merge data from multiple literature sources, resolve duplicates, and ensure consistency. This limitation restricts the broader adoption of open scientific databases and slows interdisciplinary research. 

To overcome these challenges in manual data collection and to streamline the assembly of query-specific datasets, recent advancements in Natural Language Processing (NLP) and LLMs offer promising solutions for automating literature screening and integration. Building on these advances, this work proposes an automated, scalable tool that combines keyword-based search, relevance classification using LLMs, and standardized data integration. This approach aims to reduce the manual effort required to construct custom scientific datasets from distributed literature sources, enabling researchers to generate up-to-date, task-specific databases across various scientific domains.

\subsection{NLP and LLMs for Automated Literature Screening}
NLP techniques allow us to process, interact with, and understand natural language in a structured and meaningful way. Among various NLP applications, classification tasks such as text screening are particularly valuable for organizing and evaluating large volumes of scientific literature. These techniques assist by identifying and extracting relevant content from research articles based on targeted keywords. One notable advancement in NLP is the development of transformer-based language models such as Bidirectional Encoder Representations from Transformers (BERT) \cite{bert}, which have demonstrated strong performance across a range of NLP tasks, including text classification. Building on transformers, recent advances in LLMs have further enhanced the capabilities of NLP-based classification. LLMs such as ChatGPT \cite{gpt3} by OpenAI, LLaMA \cite{llama1} by Meta, Gemma 2 by Google \cite{gemma2}, Phi 2 by Microsoft \cite{phi2}, and others have shown significant performance improvements on diverse domain-specific tasks without requiring task-specific fine-tuning. Due to the fact that LLMs are pre-trained on extensive and diverse text corpora, they can capture complex contextual relationships and perform sophisticated tasks such as literature screening with minimal additional training. In this study, we evaluate the applicability of LLMs for text classification as part of the literature screening process to support the automated assembly of open-access scientific databases and assess their effectiveness in this context.

\section{Methods}

\begin{figure}[htbp]
    \centering
    \includegraphics[scale=0.1]{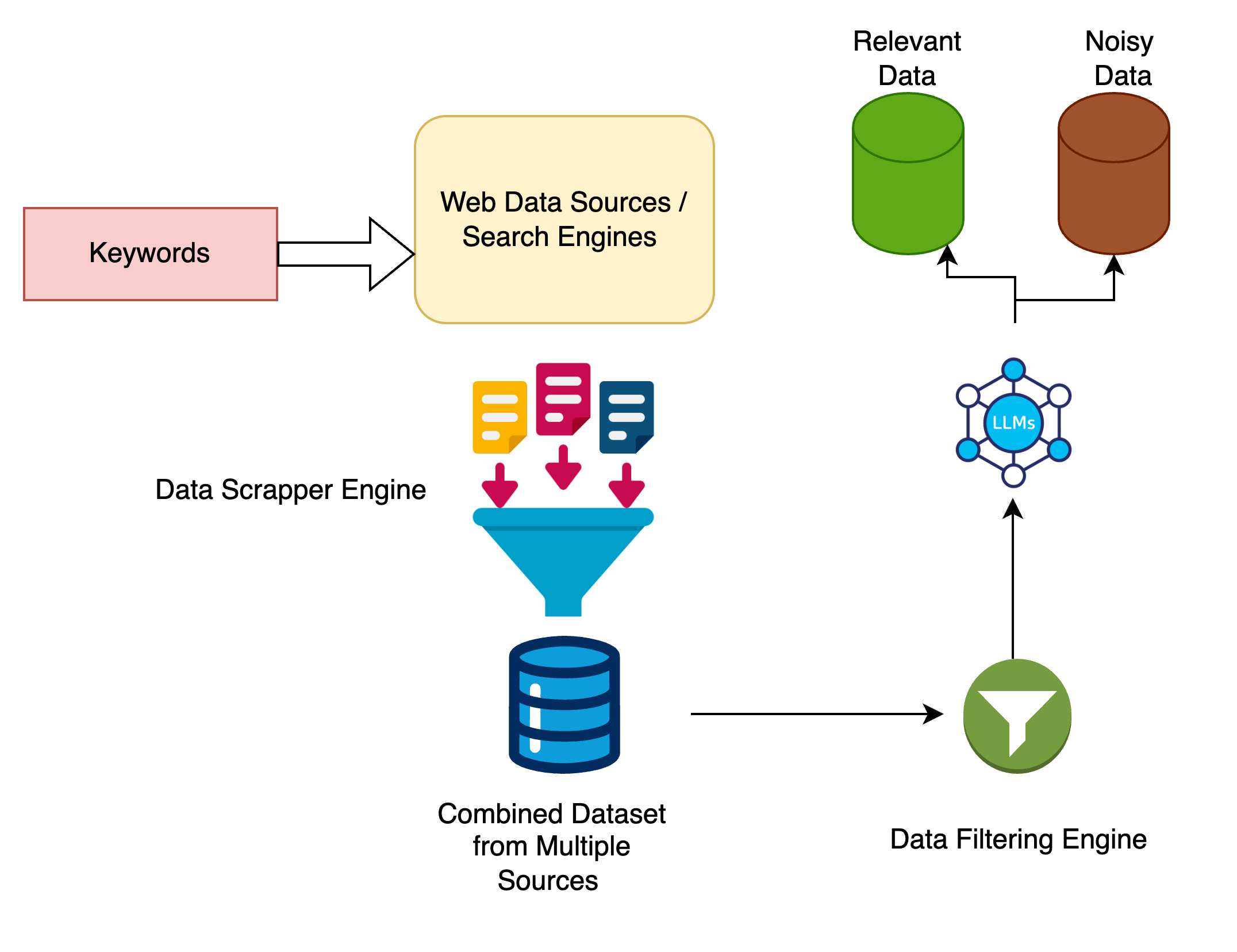}
    \caption{The figure illustrates the end-to-end pipeline for assembling a domain-specific dataset from web-based literature sources. The process begins with a set of domain-relevant keywords, which are used to query web data sources or search engines. Retrieved documents are collected by a data scraper engine, aggregating abstracts and metadata into a combined dataset from multiple sources. This raw dataset is passed through a data filtering engine, which leverages large language models (LLMs) to classify content into relevant data and noisy data. The pipeline automates and scales the screening process, enabling efficient construction of clean, task-specific datasets.}
    \label{fig:workflow}
\end{figure}

The section outlines a modular and scalable pipeline for constructing scientific datasets from multiple web resources. The pipeline consists of three primary stages: data collection, data filtering, and relevance classification. The overall architecture is illustrated in Figure \ref{fig:workflow}. To support reproducibility and enable broader use, we have released the full pipeline for the proposed tool on \href{https://github.com/NikitaGautam/AGNLP}{GithHub}.

\subsection{Data Collection}

The data collection process begins with an automated data collection system that was developed to enable large-scale literature collection across a wide range of domain-specific queries. The system can query multiple academic databases and search engines and supports keyword-driven searches across platforms, including Scopus, Web of Science, ScienceDirect, and Google Scholar. For each source, we used official APIs (where available) or automated scraping tools: 
\begin{itemize}
    \item \textbf{Scopus and ScienceDirect}: Scopus is accessed through the Elsevier Scopus API\footnote{ \url{https://dev.elsevier.com/scopus.html}}  that provides structured access to article-level metadata through their API interface. ScienceDirect is also accessed through the Elsevier ScienceDirect API\footnote{\url{https://dev.elsevier.com/sciencedirect.html}} , which supports article-level metadata retrieval. Both APIs were integrated using \href{https://pypi.org/project/pyscopus/}{Pyscopus} (a Python-based library). 
    \item \textbf{Web of Science} (WOS): WOS was queried using the  Clarivate Web of Science API\footnote{\url{https://developer.clarivate.com/apis/wos}}, which enables programmatic retrieval of search results and metadata. WOSClient\footnote{\url{https://wos.readthedocs.io/en/latest/documentation/client.html}}  is an official client library that one can use with the API Keys from WOS to get search results and article-level metadata.
    \item \textbf{Google Scholar}: Using the Scholarly\footnote{\url{https://github.com/scholarly-python-package/scholarly}} package, a custom web-scraping tool was developed to extract  publicly available metadata from Google Scholar. 
 
\end{itemize}

The data collection tool was implemented using the Flask web framework and incorporates a data scraping engine that uses API calls, when available, to retrieve metadata, titles, abstracts, document identifiers, and additional information. A key advantage of this approach is its ability to extract data from all sources concurrently, which significantly improves efficiency in terms of time and resources.
This system was applied to collect data across multiple countries, with a particular focus on Senegal, to support research on crop response to nutrient application. For example, one of the queries targeted the effect of nutrient inputs on the harvestable organ of major crops in Senegal, and was expressed as follows: \textit{Senegal AND (Nutrient OR Fertilization OR Fertilizer OR Rates OR Doses OR Nitrogen OR Phosphorus OR Potassium) AND Yield.}
Peer-reviewed articles were retrieved from Scopus, Web of Science, and ScienceDirect, while Google Scholar was used to gather both academic publications and gray literature, including institutional reports and university theses. Through this automated approach, a total of 15,825 articles were collected for the Senegal-specific search. Additional example queries for other countries are summarized in Table \ref{tab:keywords-total} (with the exception of data from Google Scholar). This generalizable framework allows for efficient, scalable, and reproducible data collection across a variety of agricultural and environmental domains.

\begin{table*}[htbp]
\scriptsize
\caption{Total number of articles collected for various domain-specific keywords}
\centering
\begin{tabular}{|p{12cm}|c|}
\hline
\textbf{Keywords} & \textbf{Total} \\
\hline
(multispectral airborne images OR drone OR UAV oR UAS OR unmanned aerial vehicle OR remotely piloted aircraft system) AND nitrogen AND yield & 77489 \\ \hline
corn OR maize AND (grain quality OR grain composition) AND (nitrogen fertilization OR water stress OR drought stress) & 63056 \\ \hline


yield AND (nitrogen fixation OR N fixation OR nitrogen from the atmosphere OR Ndfa OR nitrogen uptake OR N uptake OR seed nitrogen OR nitrogen harvest index OR NHI) AND (chickpea OR common bean OR cowpea OR faba bean OR field pea OR groundnut OR lentil OR lupins) & 4413 \\ \hline


Nitrogen dilution curve AND Nitrogen nutrition index AND Critical nitrogen concentration AND (annual ryegrass OR broomcorn millet OR cotton OR fodder beet OR hybrid ryegrass OR maize OR oat OR perennial ryegrass OR potato OR rescue grass OR rice OR sorghum OR sugarcane OR sunflower OR sweet potato OR tall fescue OR timothy grass OR wheat OR white cabbage) & 121 \\ \hline


Nigeria AND (Nutrient OR Fertilization OR Fertilizer OR Rates OR Doses OR Nitrogen OR Phosphorus OR Potassium OR Sulfur OR Sulphur) AND Yield & 5853 \\ \hline
Liberia AND (Nutrient OR Fertilization OR Fertilizer OR Rates OR Doses OR Nitrogen OR Phosphorus OR Potassium OR Sulfur OR Sulphur) AND Yield & 685 \\ \hline
Niger AND (Nutrient OR Fertilization OR Fertilizer OR Rates OR Doses OR Nitrogen OR Phosphorus OR Potassium OR Sulfur OR Sulphur) AND Yield & 5774 \\ \hline
Ghana AND (Nutrient OR Fertilization OR Fertilizer OR Rates OR Doses OR Nitrogen OR Phosphorus OR Potassium OR Sulfur OR Sulphur) AND Yield & 5527 \\ \hline
Mali AND (Nutrient OR Fertilization OR Fertilizer OR Rates OR Doses OR Nitrogen OR Phosphorus OR Potassium OR Sulfur OR Sulphur) AND Yield & 5136 \\
\hline
\end{tabular}
\label{tab:keywords-total}
\end{table*}

\subsection{Data Filtering}
\label{sec:data-filtering}
After aggregating data from multiple sources, a data cleaning step was applied to eliminate inconsistencies and ensure the collected literature dataset was error-free. The goal of this phase was to identify and remove irrelevant, redundant, and duplicate entries in order to create a unified, high-quality dataset. Initially, duplicate articles were identified by comparing document URLs, digital object identifiers (DOIs), and article titles. For sources such as Scopus and Science Direct, an additional filtering layer was applied by comparing unique \textit{Scopus\_id} to eliminate duplicate records. Although most of the retrieved content was in English, some non-English articles were included via API queries. Therefore, a language filter was implemented to detect and remove non-English entries to ensure linguistic consistency.

Table \ref{tab:deduplication_summary} illustrates an example of our data integration and filtering process by extracting data from Science Direct, Web of Science, and Scopus. We retrieved 4,999 articles from Scopus, 670 from ScienceDirect, and 638 from Web of Science. After merging Scopus and ScienceDirect, 496 duplicate entries were eliminated based on Scopus\_id, indicating substantial overlap between these two sources. After merging data from Web of Science, we applied DOI-based deduplication filtering, and observed that 277 articles had the same DOI. Furthermore, we found that due to missing or inconsistent DOI entries across the dataset, solely relying on the DOI can be a limitation. Thus, we added an additional layer of title-based filtering, which ultimately removed an additional 7 duplicates, demonstrating its effectiveness when other identifiers were incomplete. 

\begin{table}[ht]
\scriptsize
\caption{Statistics for the merging/deduplication process for Ghana-related keywords}
\centering
\begin{tabular}{|l|r|}
    \hline
    \textbf{Filtering Stage} & \textbf{Details} \\
    \hline
    \multicolumn{2}{|c|}{\textbf{Initial Record Counts}} \\
    \hline
    Scopus & 4,999 \\
    ScienceDirect & 670 \\
    Web of Science & 638 \\
    \hline
    \multicolumn{2}{|c|}{\textbf{After merging Scopus and ScienceDirect}} \\
    \hline
    Before \texttt{Scopus\_id} deduplication & 5,669 \\
    After \texttt{Scopus\_id} deduplication & 5,173 \\
    Duplicates removed (\texttt{Scopus\_id}) & 496 \\
    \hline
    \multicolumn{2}{|c|}{\textbf{After merging with WOS and deduplicating by DOI}} \\
    \hline
    Before \texttt{DOI} deduplication & 5,811 \\
    After \texttt{DOI} deduplication & 5,534 \\
    Duplicates removed (\texttt{DOI}) & 277 \\
    \hline
    \multicolumn{2}{|c|}{\textbf{After deduplicating by Title}} \\
    \hline
    Before \texttt{Title} deduplication & 5,534 \\
    After \texttt{Title} deduplication & 5,527 \\
    Duplicates removed (\texttt{Title}) & 7 \\
    \hline
    \multicolumn{2}{|c|}{\textbf{Final Record Count}} \\
    \hline
    Final number of unique records & 5,527 \\
    \hline
\end{tabular}
\label{tab:deduplication_summary}
\end{table}

\subsection{Data Classification Using LLMs}
Traditional filtering methods primarily depend on manual inspection by domain experts, who identify relevant variables through visual analysis (of titles and abstracts of the articles). This process is not only time-intensive but also costly due to the reliance on domain expertise. To mitigate this issue, we study the  ability of LLMs to understand contextual information across various domains, as LLMs require minimal supervision and have been shown to demonstrate strong capabilities in understanding textual cues in natural language text. Unlike traditional approaches that rely on domain-specific fine-tuning or expert visual inspection, LLMs provide a scalable and generalizable alternative for filtering domain-relevant literature.

In this work, we investigate the application of LLMs to automate the classification of collected articles into relevant (related) and irrelevant (unrelated) categories, thereby reducing the reliance on expert-driven, manual filtering.  Specifically, we integrate LLMs - LLaMA2-7b \cite{llama2}, Phi-2 \cite{phi2}, and Gemma-2 \cite{gemma2} - into our classification pipeline in a zero-shot setting. We perform prompt design, inference, and evaluation to extract and analyze the output from LLM. 

\begin{figure}[ht!]
    \centering
    \includegraphics[width=0.6\linewidth]{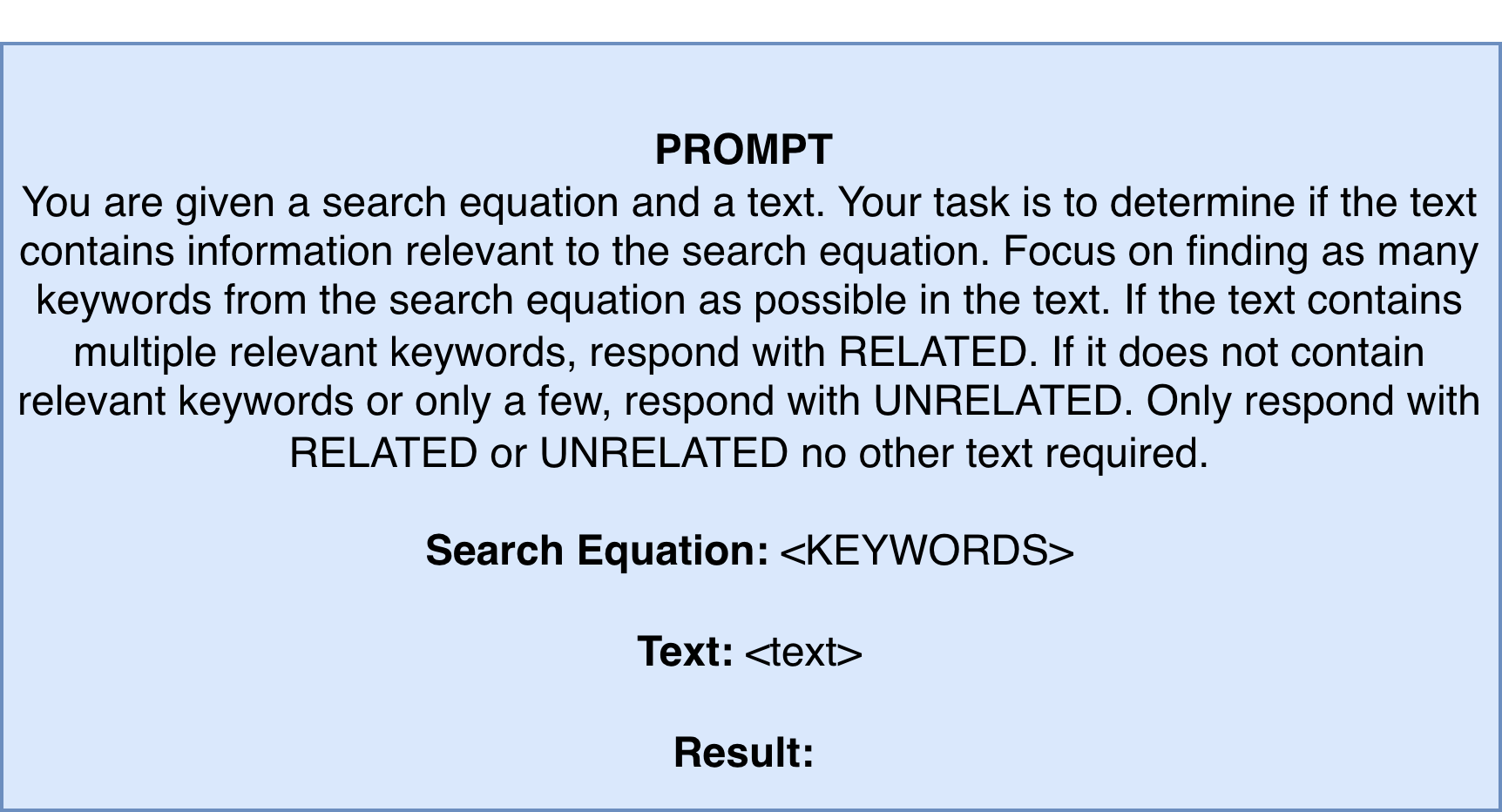}
    \caption{Prompt used for zero-shot classification using LLM models}
    \label{fig:prompt}
\end{figure}

According to domain experts, the number of keyword occurrences in a text can provide a quantitative signal regarding the relevance of the text with respect to the keywords. To account for that, keyword frequency analysis was incorporated into the zero-shot prompt used to classify texts as relevant or not relevant with LLMs. 
Figure \ref{fig:prompt}  shows the exact prompt we  used, illustrating how domain-specific keywords and contextual cues are integrated to guide the model.

The hyperparameters used to control the output of the LLM models during zero-shot learning were as follows: max\_new\_tokens was set to 32 to limit the length of the model’s response, and eos\_token\_id was defined so that the model would stop generating text at the appropriate time, do\_sample was set to True enabling sampling to allow more flexible and varied outputs, temperature was set to 0.6, and top\_p was set to 0.9. These hyperparameters yielded stable predictions during initial runs across models, and thus we used them for all models consistently for a fair comparison.

\section{Results}



\subsection{Zero-Shot Classification using LLMs}

Our domain experts defined four agriculture-related search queries, which were used to retrieve articles via the automated keyword-based pipeline. Table~\ref{tab:results} presents the performance of various LLMs in comparison with a set of articles manually filtered as relevant by domain experts, who retrieved a much smaller set of articles for each query. The \textbf{HumanRelevant} column indicates the number of relevant articles identified by experts out of the manually retrieved articles using their own methods. \textbf{ToolRetrieved} column denotes the number of articles that were automatically extracted using multiple sources by our tool. \textbf{Model} column denotes the  LLM used for classification. \textbf{ModelRelevant} refers to the number of articles classified as relevant by the LLM model from the tool-retrieved set. \textbf{HumanRelevant $\cap$ ToolRetrieved} column denotes the number of manually identified human-relevant articles that are also present in the tool-retrieved set. \textbf{HumanRelevant $-$ ToolRetrieved} column denotes the number of manually identified relevant articles that were missed by the tool. \textbf{HumanRelevant $\cap$ ModelRelevant} column denotes the number of human-relevant articles that were also predicted/classified as relevant by the model. \textbf{Overlap Accuracy} column measures the model’s ability to correctly identify relevant articles within the intersecting set of human-relevant and tool-retrieved results, calculated as the percentage of correct model-relevant items in the shared  subset, i.e. $$\textbf{\% Overlap}= 100 \times \frac{\text{HumanRelevant} \cap \text{ModelRelevant}}{\text{HumanRelevant} \cap \text{ToolRetrieved}}.$$

Evaluating the overlap accuracy for each search keyword-based query is particularly important, as it reflects the model’s performance within the shared subset of manually relevant and automatically retrieved articles. As can be seen in Table~\ref{tab:results}, all LLMs consistently achieved over 90\% accuracy across all queries. Phi-2 model had the best accuracy overall, with 100\% for 3 out of 4 queries. These results  demonstrate the effectiveness of our approach in extracting meaningful information from abstracts and accurately identifying relevant content.

When analyzing the articles labeled as relevant by the LLMs, it can be seen that the model-relevant dataset is noticeably larger than what would typically result from a manual search alone. However, there are some articles that were manually identified as relevant by the experts but not by the model, i.e., some articles were missed by the model (column {\bf HumanRelevant}$-${\bf ToolRetrieved}). An examination of those articles revealed that many of the articles not captured in the overlap were still present in alternative forms within the model-relevant dataset, although not identified by the DOI match nor by the exact title match. For example, two unmatched titles represent nitrogen in different ways (nitrogen and chemical symbol N): "determination of a critical N dilution curve for winter wheat crops" versus "determination of a critical nitrogen dilution curve for winter wheat crops". These discrepancies can often be attributed to variations in data sources, differences in how article titles are formatted across platforms, and the timing of data collection. Furthermore, search engines frequently update their indexing algorithms, which can lead to changes in search results, either excluding outdated entries or prioritizing more recent publications. These factors likely contribute to the observed gap in the overlap set. Nevertheless, overlap accuracy serves as a key indicator of the model’s effectiveness in accurately identifying relevant articles. 
The results  demonstrate the potential of LLM models to effectively replicate and potentially replace the double-blind manual approach traditionally used for data collection and filtering.

\begin{table*}[htpb]
\caption{
\scriptsize
Performance of (Llama2-7b, Gemma-2, Phi-2) models used to filter articles relevant to several domain-specific keyword-based queries, by comparison with human-labeled relevant articles.
\textbf{Keywords}: Search queries used to retrieve articles.
\textbf{HumanRelevant}: Number of articles manually identified as relevant, i.e., the number found by experts using their own methods on much smaller subsets of articles.
\textbf{ToolRetrieved}: Number of articles retrieved by the automated tool.
\textbf{HumanRelevant $\cap$ ToolRetrieved}: Number of manually identified relevant articles that are also present in the tool-retrieved set.
\textbf{HumanRelevant $-$ ToolRetrieved}: Number of manually identified relevant articles that were not retrieved by the tool.
\textbf{Model}: The LLM model used for prediction.
\textbf{ModelRelevant}: Number of articles selected as relevant by the model from the tool-retrieved set.
\textbf{HumanRelevant $\cap$ ModelRelevant}: Number of human-identified relevant articles that were also predicted as relevant by the model.
\textbf{(\% Overlap = Overlap Accuracy)}: $100 \times \frac{\text{HumanRelevant} \cap \text{ModelRelevant}}{\text{HumanRelevant} \cap \text{ToolRetrieved}}$.
}

\scriptsize{
\textbf{Multispectral}: {(multispectral airborne images OR drone OR UAV OR UAS OR unmanned aerial vehicle OR remotely piloted aircraft system) AND nitrogen AND yield}

\textbf{Corn}: {corn OR maize AND (grain quality OR grain composition) AND (nitrogen fertilization OR water stress OR drought stress)}

\textbf{N-fixation}: {yield AND (nitrogen fixation OR N fixation OR nitrogen from the atmosphere OR Ndfa OR nitrogen uptake OR N uptake OR seed nitrogen OR nitrogen harvest index OR NHI) AND (chickpea OR common bean OR cowpea OR faba bean OR field pea OR groundnut OR lentil OR lupins)}

\textbf{N-dilution}: {Nitrogen dilution curve AND Nitrogen nutrition index AND Critical nitrogen concentration AND (annual ryegrass OR broomcorn millet OR cotton OR fodder beet OR hybrid ryegrass OR maize OR oat OR perennial ryegrass OR potato OR rescue grass OR rice OR sorghum OR sugarcane OR sunflower OR sweet potato OR tall fescue OR timothy grass OR wheat OR white cabbage)}
}

\centering
\renewcommand{\arraystretch}{1.25}
\setlength{\tabcolsep}{3pt}

\resizebox{\textwidth}{!}{%
\begin{tabular}{@{}lccccccccc@{}}
\toprule
\textbf{Keywords} &
\textbf{Human} &
\textbf{Tool} &
\textbf{HumanRelevant} &
\textbf{HumanRelevant} &
\textbf{Model} &
\textbf{Model} &
\textbf{HumanRelevant} &
\textbf{(\%)} \\
& \textbf{Relevant} &
\textbf{Retrieved} &
\textbf{$\cap$ ToolRetrieved} &
\textbf{$-$ ToolRetrieved} &
 &
\textbf{Relevant} &
\textbf{$\cap$ ModelRelevant} &
\textbf{Overlap} \\
\midrule

\multirow{3}{*}{Multispectral} 
& \multirow{3}{*}{41} & \multirow{3}{*}{77489} & \multirow{3}{*}{36} & \multirow{3}{*}{5} & Llama2-7b & 47666 & 33 & 91.67 \\
&  &  &  &  & Gemma-2 & 68063 & 33 & 91.67 \\
&  &  &  &  & Phi-2 & 78681 & 36 & 100 \\ 
\midrule

\multirow{3}{*}{Corn} 
& \multirow{3}{*}{21} & \multirow{3}{*}{63056} & \multirow{3}{*}{13} & \multirow{3}{*}{8} & Llama2-7b & 48752 & 12 & 92.31 \\
&  &  &  &  & Gemma-2 & 57569 & 12 & 92.31 \\
&  &  &  &  & Phi-2 & 62460 & 13 & 100 \\ 
\midrule

\multirow{3}{*}{N-fixation} 
& \multirow{3}{*}{83} & \multirow{3}{*}{4413} & \multirow{3}{*}{46} & \multirow{3}{*}{37} & Llama2-7b & 3644 & 43 & 93.48 \\
&  &  &  &  & Gemma-2 & 3991 & 39 & 84.78 \\
&  &  &  &  & Phi-2 & 4282 & 43 & 93.48 \\ 
\midrule

\multirow{3}{*}{N-dilution} 
& \multirow{3}{*}{34} & \multirow{3}{*}{121} & \multirow{3}{*}{14} & \multirow{3}{*}{20} & Llama2-7b & 120 & 14 & 100 \\
&  &  &  &  & Gemma-2 & 109 & 13 & 92.86 \\
&  &  &  &  & Phi-2 & 121 & 14 & 100 \\ 

\bottomrule
\end{tabular}
}

\label{tab:results}
\end{table*}

\section{Abstract Filtering Tool}

\begin{figure*}[htbp]
    \centering
    \includegraphics[width=1\linewidth]{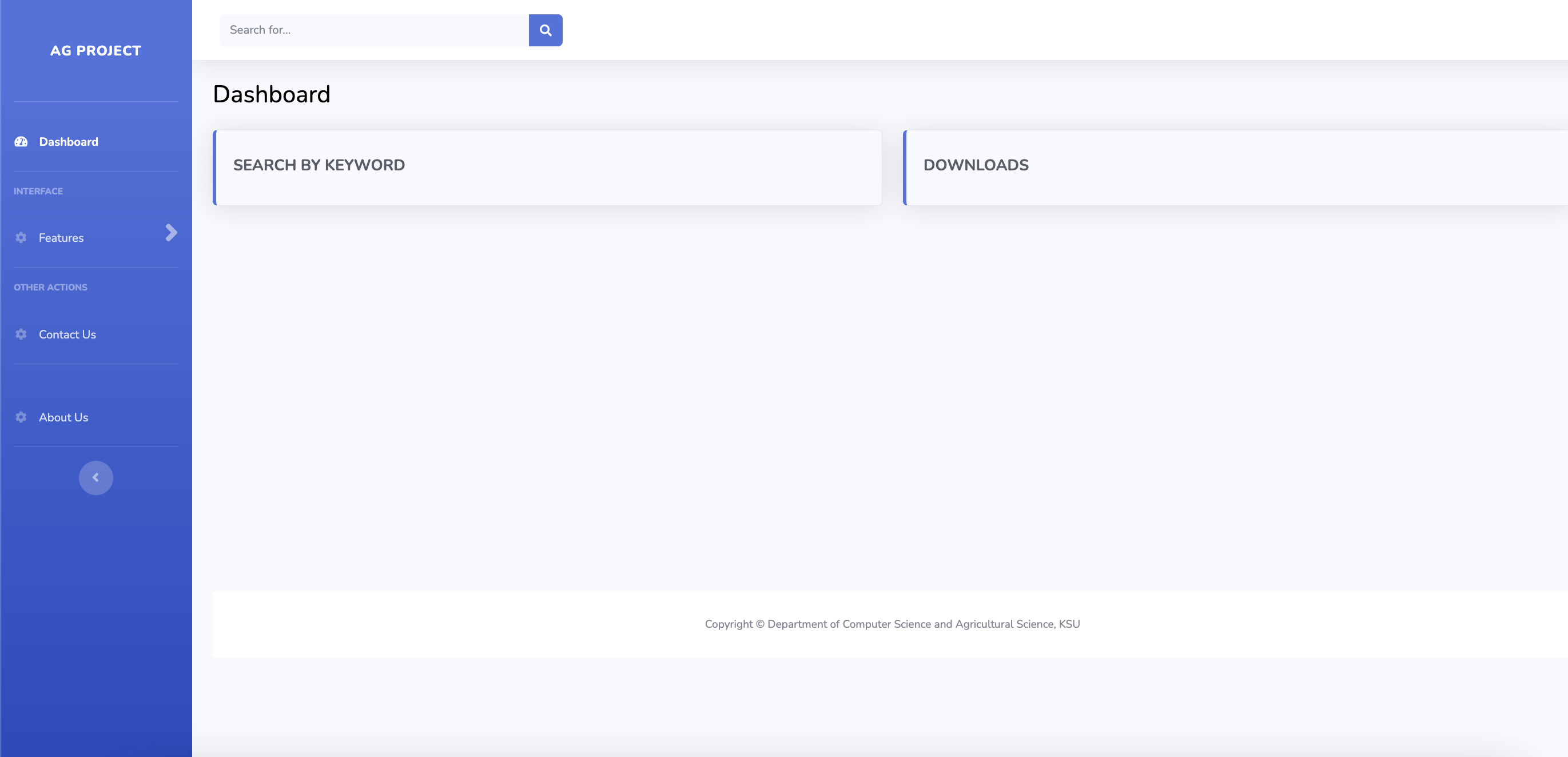}
    \caption{Dashboard of Abstract Filtering Tool}
    \label{fig:dashboard}
\end{figure*}

To enable scalable and automated data collection and classification, we developed a web-based tool using the Flask framework. This tool comprises several integrated components, including a data scraping module that retrieves literature from academic sources such as Scopus, Web of Science, Science Direct, and Google Scholar. API calls are used to extract key metadata fields, such as titles, abstracts, authors, and document identifiers, as well as other search engine-specific information, from each database or search engine. Figure \ref{fig:dashboard} shows the dashboard of our web-based application, which allows users to efficiently search for data using specific keywords in multiple sources or to access a download page to retrieve previously collected datasets. 
Data collection is executed in parallel across multiple sources, ensuring faster retrieval and improved accessibility.

\begin{figure*}[htbp]
    \centering
    \includegraphics[width=1\linewidth]{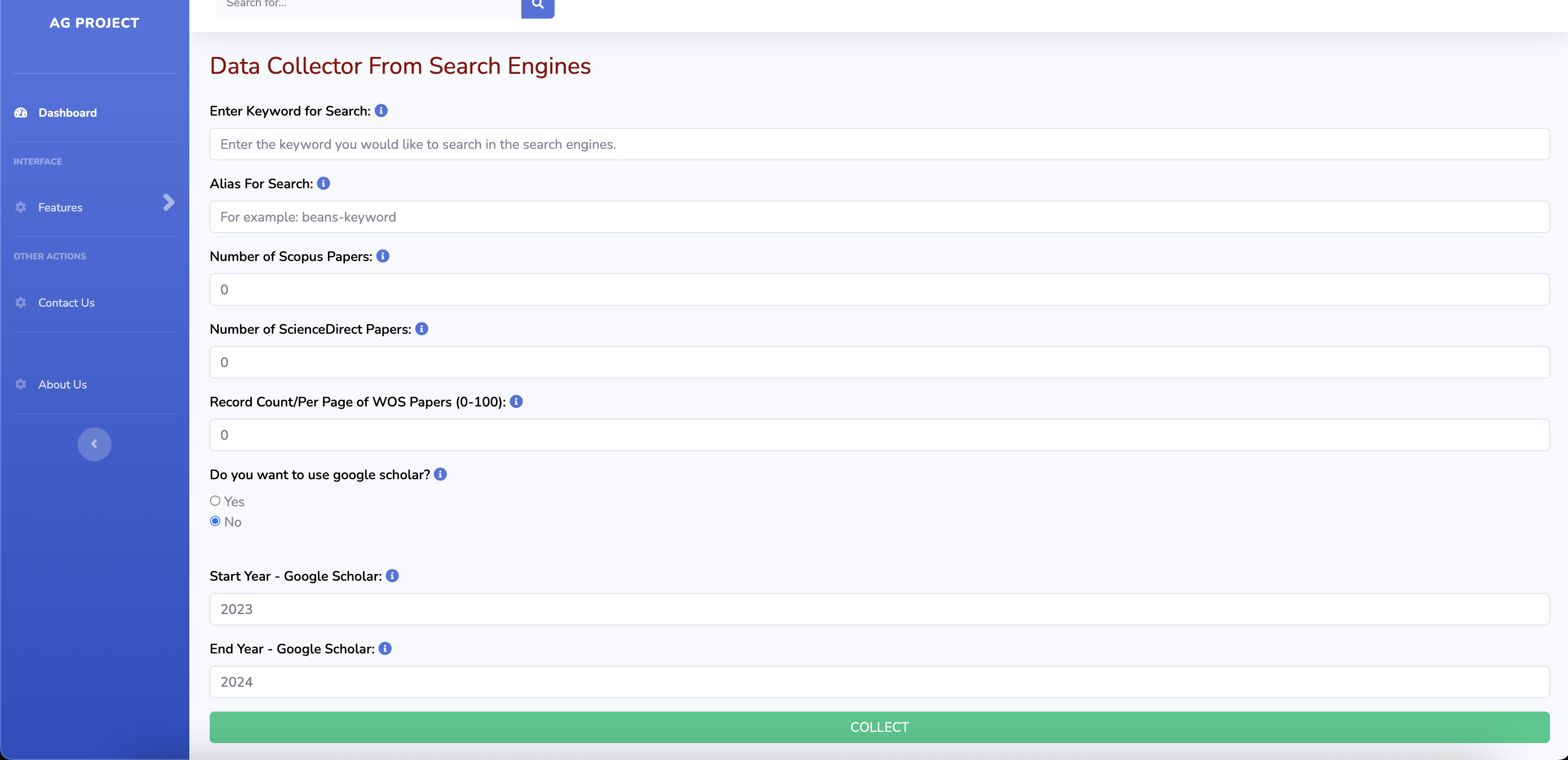}
    \caption{Data Collection Page of Abstract Filtering Tool}
    \label{fig:datacollection}
\end{figure*}


The simple input form serves as a one-stop interface for initiating keyword-based searches across multiple data sources/search engines, as shown in Figure \ref{fig:datacollection}. Users can specify a custom alias for each search, which is later used to label the corresponding dataset on the downloads page. To ensure uniqueness, each alias must be distinct from all others. For Scopus and Science Direct, users can define the number of articles to retrieve, with a maximum limit of 5,000 papers per source. For Web of Science (WOS), data is retrieved by page, and users can enter a value between 0 and 100 to determine the number of pages to scrape per API request. This ensures that all available articles within those pages are extracted. An additional option is available to include Google Scholar as a data source. This option is set to \textit{No}, by default, due to the platform’s restrictions on automated scraping and the absence of abstract metadata, which limits the effectiveness of downstream classification performance. If users select to include Google Scholar, the tool will use article titles for classification instead of abstracts. Users can also specify a range of publication years, with up to 1000 articles collected per year. Once the user clicks the \textit{Collect} button, the system starts to retrieve data from all selected sources concurrently, enabling efficient and scalable data collection.

After data collection, the tool automatically performs a series of preprocessing steps, including removing duplicate entries, filtering out non-English content, and validating metadata. The cleaned data is then passed to a classification layer/module powered by an LLM in the background, which categorizes articles as relevant or irrelevant using domain-specific prompts related to the current search keywords. Finally, users can download the collected data in an open-access CSV format, which supports transparency and facilitates easy reuse.

The tool was designed to leverage parallel processing, enabling the simultaneous querying of multiple data sources and significantly improving both speed and efficiency. It also incorporates features that enable flexible adaptation across various research domains. By automating tasks that would otherwise require extensive manual effort and time from domain experts, the tool provides a scalable and effective solution applicable far beyond its original use case in agricultural literature. Although broad, cross-disciplinary open data initiatives have struggled to achieve widespread adoption within the scientific community, domain-specific platforms like this one hold greater potential for promoting effective data sharing and improving accessibility.

\section{Conclusion and Future Work}
The paper presents a methodology that can be generalized and scaled for the automated collection, classification, and organization of scientific literature using LLMs. The proposed method enables the efficient collection of large volumes of data from multiple data sources/search engines. It achieves strong classification performance (i.e., an accuracy of over 90\% overlap between data selected by domain experts and data selected by the LLM model, without requiring expert supervision). The tool introduced in the paper supports the creation of open-access, domain-specific databases and promotes reproducible research by providing downloadable datasets.

One limitation of the tool is the restrictions on data scraping from platforms like Google Scholar. Additionally, classification accuracy may vary depending on the clarity of the abstracts, the relevance of the keywords used in the search query, and the LLMs employed for zero-shot learning. Future work might include using few-shot learning to improve classification accuracy. Although the proposed method is designed to be domain-agnostic, incorporating additional domain-specific data sources that can be dynamically integrated into the search process is an important direction for future work.

\section*{Acknowledgments}

Research reported in this publication was supported by USDA-NRCS grant (NR233 A750011G001). The research was also partially supported by the Cognitive and Neurobiological Approaches to Plasticity (CNAP) Center of Biomedical Research Excellence (COBRE) of the National Institutes of Health under grant number P20GM113109. The content is solely the responsibility of the authors and does not necessarily represent the official views of the National Institutes of Health and USDA-NRCS.

\bibliographystyle{splncs04}
\bibliography{ag-bib}

\end{document}